\documentclass{PoS}
\usepackage{url}

\title{Pre-Big Bang, vacuum and noncyclic cosmologies}

\ShortTitle{Noncyclic pre-Big Bang}

\author{\speaker{Luis Gonzalez-Mestres}\\
        LAPP, Universit\'e de Savoie, CNRS/IN2P3\\ 
        B.P. 110, 74941 Annecy-le-Vieux Cedex, France\\
        E-mail: \email{lgm\_sci@yahoo.fr}}

\abstract{WMAP and Planck open the way to unprecedented Big Bang phenomenology, potentially allowing to test the standard Big Bang model as well as less conventional approaches including pre-Big Bang cosmologies. An illustration is provided by the recent claim (Gurzadyan et al. \cite{Gurzadyan}) that the cosmological sky would be a weakly random one with mostly regular signal. This work has been followed by an interesting and useful debate. Whatever the conclusion, it appears that a detailed study of WMAP and Planck data can have significant implications for pre-Big Bang theories. Not only for cyclic cosmologies following the analysis recently proposed by Gurzadyan and Penrose \cite{GurzadyanPenrose}, but also for noncyclic approaches incorporating a new fundamental scale beyond the Planck scale and, possibly, new ultimate constituents of matter with unconventional basic properties as compared to standard particles \cite{Gonzalez-Mestres1,Gonzalez-Mestres2}. Alternatives to standard physics can be considered from a cosmological point of view concerning vacuum structure, the nature of space-time, the origin and evolution of our Universe, the validity of quantum field theory and conventional symmetries, solutions to the cosmological constant problem, inflationary scenarios, dark matter and dark energy, the interpretation of string-like theories... Lorentz-like symmetries for the properties of matter (standard or superbradyonic \cite{Gonzalez-Mestres3,Gonzalez-Mestres4}) can then be naturally stable space-time configurations resulting from general cosmological scenarios that incorporate physics beyond the Planck scale and describe the formation and evolution of the present vacuum. But an even more primordial question seems to be that of the origin of half-integer spins, that cannot be generated through orbital angular momentum in the usual real space-time. It turns out that the use of a spinorial space-time \cite{Gonzalez-Mestres5,Gonzalez-Mestres6} with two complex coordinates instead of the conventional four real ones presents several attractive features. Taking the cosmic time to be the modulus of a SU(2) spinor leads by purely geometric means to a naturally expanding universe \cite{Gonzalez-Mestres6,Gonzalez-Mestres7}, with a ratio between cosmic relative velocities and distances equal to the inverse of the age of the Universe. No reference to standard matter, hidden fields, gravitation or relativity is required to get such a result that looks quite reasonable from an observational point of view. We discuss basic ideas and phenomenological issues for noncyclic pre-Big Bang cosmologies in the present context.} 

\FullConference{The 2011 Europhysics Conference on High Energy Physics-HEP 2011,\\
		July 21-27, 2011\\
		Grenoble, Rh\^one-Alpes, France}

\begin{document}
\section{Beyond Big Bang and Planck scale}

Eighty years after the Big Bang hypothesis formulated by Georges Lema\^itre \cite{Lemaitre1}, WMAP and Planck may allow to explore the origin of the Universe, as well as the structure of matter and space-time, beyond the primeval quanta he postulated. Issues like the possible lack of randomness of cosmic microwave background (CMB) \cite{Gurzadyan} are relevant not only for conformal cyclic cosmologies (CCC) \cite{GurzadyanPenrose}. They can also be crucial for pre-Big Bang patterns with a possible new era before the Big Bang, new properties of matter \cite{Gonzalez-Mestres1,Gonzalez-Mestres2} and a new space-time structure \cite{Gonzalez-Mestres5,Gonzalez-Mestres6}. These approaches naturally lead to alternatives \cite{Gonzalez-Mestres2,Gonzalez-Mestres4} to standard cosmology, inflation, dark matter and dark energy.  

A pre-Big Bang scenario based on string models also exists {\cite{Veneziano}, but strings can result from an underlying composite structure \cite{Gonzalez-Mestres4,Gonzalez-Mestres7} and no longer make sense below some distance scale. 

Contrary to early models that used preons as mere building blocks with similar properties to those of standard particles, the superbradyon hypothesis proposed in 1995 \cite{Gonzalez-Mestres3,Gonzalez-Mestres4} assumes that the critical speed $c_s$ of the new preonic constituents is much larger than the speed of light $c$, just as $c$ is about a million times the speed of sound. The physical vacuum is then a material medium ultimately made of the actual fundamental matter (the superbradyons), where conventional particles can exist as excitations similar to phonons, solitons... The choice of a Lorentz metric for superbradyons with $c_s$ as the critical speed appears natural, as other space-time metrics are expected to produce vacuum instabilities \cite{Gonzalez-Mestres1}. The existence of a preferred reference frame (the "vacuum rest frame", VRF) is a basic and permanent ingredient of such new physics and cosmology.  

If the vacuum is made of superbradyonic matter, standard gauge theories and conventional symmetries (including Lorentz symmetry) will provide only a sectorial low-energy limit. The Higgs boson and the zero modes of bosonic harmonic oscillators do not need to be permanently materialized in vacuum. The cosmological constant problem may then be naturally solved. Quantum field theory, including the calculation of Feynman diagrams, is expected to undergo modifications at very high energy \cite{Gonzalez-Mestres1,Gonzalez-Mestres2}. Quantum mechanics and other currently admitted principles of Physics will not necessarily hold beyond the Planck scale or even before reaching it \cite{Gonzalez-Mestres6}. 

It is usually believed that standard symmetries of Particle Physics should look more and more exact as the energy scale considered increases and masses can be neglected. But it may actually happen that, above some transition energy below Planck scale, the situation changes and tracks from new physics generated at the Planck scale and beyond become increasingly important \cite{Gonzalez-Mestres6,Gonzalez-Mestres10}.  

If free superbradyons can exist in our Universe with speeds larger than $c$, they would spontaneously emit "Cherenkow" radiation in the form of standard particles. Remnant superbradyons with a speed close to $c$ can then form a cosmological sea and a new dark matter component \cite{Gonzalez-Mestres2,Gonzalez-Mestres6}.  

In this context, two kinds of pre-Big Bang scenarios can be considered : i) a pre-Universe made of the actual ultimate constituents of matter and ruled by their own dynamical laws ; ii) an initial singularity followed by a process generating the "primeval quanta" \cite{Lemaitre1} of the Big Bang. The initial singularity in ii) can actually correspond to a nucleation inside the pre-Universe of i). The Big Bang itself can be replaced by a superbradyon era \cite{Gonzalez-Mestres4,Gonzalez-Mestres7} with a transition to standard matter at a lower temperature, thus providing a direct alternative to inflation \cite{Gonzalez-Mestres4,Gonzalez-Mestres8}. The space-time geometry is crucial for the study of pre-Big Bang patterns and can lead to unexpected modifications of conventional views concerning the origin and evolution of our Universe.

\section{A spinorial space-time}

As spin-1/2 particles exist in our Universe, the most natural description of space-time would be a spinorial one \cite{Gonzalez-Mestres5,Gonzalez-Mestres6} with, at least, a SU(2) symmetry group \cite{Gonzalez-Mestres9}. With these minimal hypotheses, taking a preferred reference frame as suggested by cosmological data and required if superbradyons exist, a cosmic time can be defined. Given a spinor $\xi $, and considering the positive SU(2) scalar $\mid \xi \mid ^2$ $=$ $\xi ^\dagger \xi $ where the dagger stands for hermitic conjugate, the cosmic time would be $t~=~\mid \xi \mid$ and the associated space given by the $S^3$ hypersphere $\mid \xi \mid~=~t$. Then, if $\xi _0$ is the observer position on the $\mid \xi \mid $ = $t_0$ hypersphere, space translations correspond to SU(2) transformations acting on the spinor space, i.e. $\xi ~=~ U ~\xi _0$ where $U~=~exp~(i/2~~t_0^{-1}~{\vec \sigma }.{\vec {\mathbf x}})~ \equiv U ({\vec {\mathbf x}})$, and ${\vec \sigma }$ is the vector formed by the Pauli matrices. The vector ${\vec {\mathbf x}}$ is the spatial position of $\xi $ with respect to $\xi _0$ at constant time $t_0$, and is different from the spinorial position $\xi ~- ~\xi _0$. Space rotations are obtained as SU(2) transformations acting on the spatial position vector with respect to a fixed point $\xi _0$. The origin of our time can be associated to the point $\xi ~= ~0$. This leads to a naturally expanding Universe where cosmological comoving frames would correspond to straight lines crossing the origin $\xi ~= ~0$.
 
Such a geometry automatically yields the well-known relation between relative velocities and distances at cosmic scale for comoving frames, usually called Hubble's law but actually first formulated by Georges Lema\^itre \cite{Lemaitre2,Hubble}. In our spinorial approach to space-time, if $\theta $ is a constant angular distance between two cosmological comoving frames, the $S^3$ spatial distance $d$ between the two corresponding points on the $\mid \xi \mid ~=~t$ hypersphere will be $d ~= ~\theta ~t$. The ratio between relative velocities and distances is then given by the inverse of the age of the Universe. This value is in reasonable agreement with present observations while matter, standard relativity, gravitation and specific space units have not yet been introduced in our description of space-time. It is therefore tempting to conjecture that the usually postulated dark energy is not required to explain the observed acceleration of the expansion of the Universe. Instead, gravitational and other standard effects can possibly account for past fluctuations of the velocity/distance ratio \cite{Gonzalez-Mestres6,Gonzalez-Mestres7}. 

Although the spinorial relative position $\xi ~- ~\xi _0$ defined above corresponds to a path through past times violating standard causality, and the production of half-integer orbital angular momenta in experiments has never been reported, it may happen that such a position spinor makes sense at very small distance scales. It would then be possible to generate the spin 1/2 as an actual internal orbital momentum in an underlying composite picture of quarks and leptons possibly linked to a pre-Big Bang cosmology. Then, the spinorial space-time would also be crucial for our understanding of the ultimate structure of matter beyond conventional quantum field theory. 

Contrary to the conventional mathematical structure of the Poincar\'e group, the spinorial space-time discussed here incorporates space translations and rotations in a single compact group. This is radically different from the assumptions that led to the Coleman-Madula theorem in 1967 \cite{ColemanMandula}. Therefore, the present SU(2) approach to space-time and its natural SL(2,C) extension can open the way to a new unification between space-time and internal symmetries \cite{Gonzalez-Mestres6}. Such a new unified symmetry may in turn provide indirect checks of the spinorial space-time pattern suggested. 

This SU(2) description of space-time considered here is also close to a SO(4) approach where, instead of being imaginary, the cosmic time would be given by the modulus of a four-vector \cite{Gonzalez-Mestres6}. To date, cosmological data have not excluded a $S^3$ hyperspherical Universe. 

\section{Conclusion}

Perhaps the development of Particle Physics and Cosmology is just beginning, and the really fundamental objects and phenomena remain to be discovered and studied. Together with direct cosmological observations, ultra-high energy cosmic rays \cite{Gonzalez-Mestres10} can help to explore new physics.

\end{document}